\documentclass[%
twocolumn,
noeprint,
superscriptaddress,
nolongbibliography,
 amsmath,amssymb,
 aps,
]{revtex4-2}

\usepackage{graphicx}
\usepackage{dcolumn}
\usepackage{bm}
\usepackage{hyperref}
\usepackage{threeparttable}
\usepackage{enumerate}
\usepackage{bbding}
\usepackage{color}
\usepackage{multirow}
\usepackage{ulem} 
\usepackage{xcolor} 

\renewcommand{\sout}{\bgroup \color{red} \ULdepth=-.5ex \ULset}

\begin{document}


\title{Bayesian model averaging for nuclear symmetry energy from effective proton-neutron chemical potential difference of neutron-rich nuclei}

\author{Mengying Qiu}
\affiliation{Sino-French Institute of Nuclear Engineering and Technology, Sun Yat-Sen
University, Zhuhai 519082, China}

\author{Bao-Jun Cai}
\email[Corresponding author:]{bjcai87@gmail.com}
\affiliation{Quantum Machine Learning Laboratory, Shadow Creator Inc., Shanghai 201208, China}

\author{Lie-Wen Chen}
\email{lwchen@sjtu.edu.cn}
\affiliation{School of Physics and Astronomy and Shanghai Key Laboratory for Particle
Physics and Cosmology, Shanghai Jiao Tong University, Shanghai 200240, China}

\author{Cen-Xi Yuan}
\email{yuancx@mail.sysu.edu.cn}
\affiliation{Sino-French Institute of Nuclear Engineering and Technology, Sun Yat-Sen
University, Zhuhai 519082, China}

\author{Zhen Zhang}
\email[Corresponding author:]{zhangzh275@mail.sysu.edu.cn}
\affiliation{Sino-French Institute of Nuclear Engineering and Technology, Sun Yat-Sen
University, Zhuhai 519082, China}


\date{\today}

\begin{abstract}

The data-driven Bayesian model averaging is a
rigorous statistical approach to combining multiple models for a unified prediction. Compared with the individual model, it provides more reliable information, especially for problems involving apparent model dependence. 
In this work, within both the non-relativistic Skyrme energy density functional and the nonlinear relativistic mean field model, 
the effective proton-neutron chemical potential difference $\Delta \mu^*_{\rm{pn}}$ of neutron-rich nuclei is  found to be strongly sensitive to the symmetry energy $E_{\rm{sym}}(\rho)$ around  $2\rho_0/3$,  with $\rho_0$ being the nuclear saturation density. 
Given discrepancies on the $\Delta \mu^*_{\rm{pn}}$-$E_{\rm{sym}}(2\rho_0/3)$ correlations between the two models, we carry out  a Bayesian model averaging analysis based on Gaussian process emulators to extract the symmetry energy around $2\rho_0/3$ from 
the measured $\Delta \mu^*_{\rm{pn}}$  of 5 doubly magic nuclei $^{48}$Ca, $^{68}$Ni, $^{88}$Sr, $^{132}$Sn and $^{208}$Pb. Specifically,  the  $E_{\mathrm{sym}}(2\rho_0/3)$ is inferred to be $E_{\mathrm{sym}}(2\rho_0/3) = 25.6_{-1.3}^{+1.4}\,\mathrm{MeV}$ at $1\sigma$ confidence level. The obtained constraints on the $E_{\mathrm{sym}}(\rho)$ around $2\rho_0/3$ agree well with microscopic predictions and results from other isovector indicators.

\end{abstract}

\maketitle



As nuclear symmetry energy $E_{\mathrm{sym}}(\rho)$ governs the isospin-dependence of nuclear matter equation of state (EOS), its density dependence has long been a subject of intense study in nuclear physics and astrophysics \cite{Baran2005,Steiner2005b,Lattimer:2006xb,Li:2008gp}.  Thanks to the rapid development in terrestrial nuclear experiments and astrophysical observations, great progress has been achieved recently in exploring the density dependence of the symmetry energy \cite{Baldo:2016jhp,Oertel:2016bki,Roca-Maza:2018ujj,Li:2021thg}. However, various challenges emerge concurrently. For example,  very recently, the PREX-2 and CREX collaborations have respectively measured  the neutron skin thickness (one of the most important probes of the density dependence of symmetry energy) in $^{208}$Pb and $^{48}$Ca with minimum strong-interaction model dependence \cite{PREX:2021umo, CREX:2022kgg}. Unfortunately, the PREX-2 and CREX results exhibit a significant tension, which challenges current theoretical models and our understanding on  the density dependence of the symmetry energy \cite{Reed:2021nqk, Piekarewicz:2021jte,Reinhard:2021utv,
Reinhard:2022inh,Yuksel:2022umn, Zhang:2022bni}. 
The accurate determination of the density dependence of symmetry energy 
remains an open challenge. 

An important issue in the study of nuclear symmetry energy is the model dependence. Although there are some isovector indicators, like the neutron skin thickness \cite{Brown:2000pd,Typel:2001lcw,Furnstahl:2001un, Chen:2005ti} and the electric dipole polarizability \cite{Roca-Maza:2013mla, Zhang:2014yfa,  Zhang:2015ava,Roca-Maza:2015eza}, 
whose correlations to the symmetry energy have been proved to be model independent, it is not the case for most of other probes. Analyses with different models on the same observable may result in different and even contradictory conclusions, e.g., the case of charged pion ratio \cite{Xiao:2008vm,Feng:2009am, Xie:2013np,Hong:2013yva,Cozma:2016qej,Zhang:2017mps,SpiRIT:2020sfn, Yong:2021nwn}. Even for the widely used nuclear energy density functional theory (EDF), such as the Skyrme EDF and the nonlinear relativistic mean field (RMF) model, there exist systematic discrepancies among their predictions for the saturation properties of nuclear matter \cite{Chen:2007,Chen:2009wv, Dutra:2012mb,Dutra:2014qga,Sun:2023xkg}. The model dependence issue hinders the reliable uncertainty quantification of the density dependence of symmetry energy. 

As a powerful statistical inference method, the Bayesian approach  
has been extensively used to extract physics information from experiments and observations. Within the Bayesian framework, the model dependence issue can be addressed via the Bayesian model averaging(BMA), which mixes predictions of multiple models according to their capabilities of reproducing given data~\cite{RevModPhys.83.943,Kejzlar_2020,Jay:2020jkz}. In nuclear physics studies, the BMA has been successfully used to 
predict the neutron drip line using various nuclear EDFs~\cite{Neufcourt:2019qvd} and to extract the shear and bulk viscosities of quark-gluon plasma from measurements in heavy-ion collision~\cite{JETSCAPE:2020shq}. It has also been proposed to quantify the inter-model uncertainties in some important nuclear physics problems, such as 
the nuclear matrix elements governing neutrinoless double-beta decay~\cite{Cirigliano:2022rmf}.

In this work, within both the non-relativisitic Skyrme EDF and the nonlinear RMF model, we find the effective proton-neutron chemical potential difference $\Delta \mu^*_{\rm{pn}}$ is strongly correlated with the symmetry energy around a subsaturation density of $2\rho_0/3$, with $\rho_0$ being the saturation density. However, there exists systematic discrepancy between the correlations predicted by the two models. Based on Gaussian process (GP) emulators trained respectively with a number of Skyrme EDFs and nonlinear RMF parameter sets, the symmetry energy around $2\rho_0/3$
are extracted from Bayesian analyses of the $\Delta \mu^*_{\rm{pn}}$ in $^{48}\text{Ca}$, $^{68}\text{Ni}$, $^{88}\text{Sr}$, $^{132}\text{Sn}$ and $^{208}\text{Pb}$. Further Bayesian model averaging over the results from the two models gives more reliable constraints on the symmetry energy around $2\rho_0/3$ by taking into account the model dependence.

The neutron and proton chemical potential in a nucleus with $N$ neutrons and $Z$ protons can be  approximated respectively to be 
\begin{eqnarray}
  \mu_{\rm{n}} &=& \frac{\partial B(N,Z)}{\partial N}\approx \frac{B(N+2,Z)-B(N-2,Z)}{4}, \\
  \mu_{\rm{p}} &=& \frac{\partial B(N,Z)}{\partial Z}\approx \frac{B(N,Z+2)-B(N,Z-2)}{4},
\end{eqnarray}
with $B(N,Z)$ being the binding energy of the nucleus. {Although single nucleon separation energy can also be used to approximate the nucleon chemical potential, it suffers from pairing effects, and strong shell effects at magic numbers. Comparing with the single nucleon separation energy, the adopted approximation for $\mu_{\rm{n(p)}}$ is relatively free of pairing and shell effects.  } 
The effective proton-neutron chemical potential difference is then defined by
\begin{eqnarray}\label{eq:Dmu}
\Delta \mu^*_{\rm{pn}} &=&  \frac{1}{4}\left[ B(N,Z+2)-B(N,Z-2)\right. \notag \\ 
&& \left. -B(N+2,Z)+B(N-2,Z)\right].
\end{eqnarray}

The sensitivity of the $\Delta \mu^*_{\rm{pn}}$   to the symmetry energy can be expected from the semi-empirical mass formula 
$
B(N, Z)=a_{\rm{v}} A-a_{\rm{s}} A^{2 / 3}-a_{\rm{c}} \frac{Z^2}{A^{1 / 3}}-a_{\mathrm{sym}} I^2 A+E_{\rm{mic}},
$
where $A=N+Z$  is the mass number, $I = (N-Z)/A$ is the isospin asymmetry, $a_{\rm{v}}$, $a_{\rm{s}}$, $a_{\rm{c}}$ and $a_{\mathrm{sym}}$ are the coefficients of volume, surface, Coulomb and asymmetry terms, respectively, and the microscopic correction $E_{\rm{mic}}$ includes pairing and shell effects. Considering $2/A$ as a small quantity, one can derive  
$\Delta \mu^*_{\rm{pn}} 
\simeq -2 a_{\rm{c}} \frac{Z}{A^{1 / 3}}+  4 a_{\mathrm{sym}}I $, 
which suggests that the $\Delta \mu^*_{\rm{pn}}$ is essentially determined by the symmetry energy of finite nuclei, since the Coulomb coefficient $a_{\rm{c}} \simeq 0.71$ MeV~\cite{Wangning:2011} is well determined. Given the well-confirmed empirical relationship $a_{\mathrm{sym}} \simeq E_{\mathrm{sym}}(\rho_A)$ at a subsaturation reference density $\rho_A$ in numerous studies~\cite{Centelles:2009,Chen:2011,Wang:2013},  one can expect that there exist  strong correlations between $\Delta \mu^*_{\rm{pn}}$  of neutron-rich doubly magic nuclei and $E_{\mathrm{sym}}(\rho)$ at subsaturation densities.
We note that the $\Delta \mu^*_{\rm{pn}}$  is similar to the neutron-proton Fermi energy difference proposed in Ref.~\cite{Wang:2013}. Nevertheless, the $\Delta \mu^*_{\rm{pn}}$ can be directly calculated by nuclear energy density functionals and deduced from experimental measured nuclear masses.

\begin{figure}[ht]
  \includegraphics[width=0.465\textwidth]{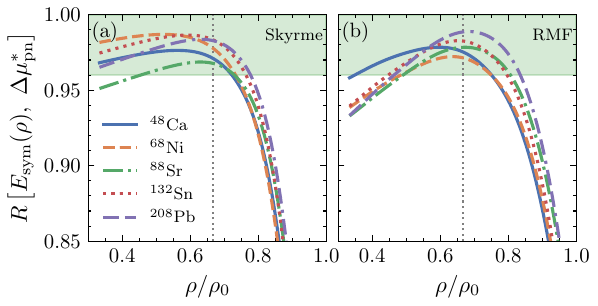}
  \caption{Pearson correlation coefficients between the $E_{\text{sym}}(\rho)$ and the $\Delta \mu^*_{\rm{pn}}$ in $^{48}\text{Ca},~^{68}\text{Ni},~^{88}\text{Sr},~^{132}\text{Sn}$ and $^{208}\text{Pb}$ as functions of nucleon density $\rho$ predicted by the sampled 50 Skyrme EDFs (left window) and nonlinear RMF parameter sets (right window). The vertical grey dotted line indicates $\rho = 2\rho_0/3$. The green shaded region represents $R>0.96$. \label{fig:CorCoeff}}
  \end{figure}

We further study the correlation between the effective proton-neutron chemical potential difference and the symmetry energy at subsaturation densities using a number of non-relativistic Skyrme and covariant EDFs. 
The Skyrme EDF is taken to be the conventional standard form which contains 9 Skyrme parameters of $x_0-x_3$, $t_0-t_3$ and $\alpha$, and a spin-orbit parameter $W_0$\cite{Chabanat:1997qh}. The 9 Skyrme parameters $t_0- t_3$, $x_0- x_3$ and $\alpha$ can be expressed in terms of 9 pseudo-observables in nuclear matter:  the nuclear matter saturation density $\rho_0$, the energy per particle of symmetric nuclear matter $E_0(\rho_0)$, the incompressibility $K_0$, the isoscalar effective mass $m_{s,0}^{\ast}$, the isovector effective mass $m_{v,0}^{\ast}$, the gradient coefficient $G_S$, the symmetry-gradient coefficient $G_{V}$, and the magnitude $E_{\mathrm{sym}}(\rho_0)$ and density slope $L$ of the nuclear symmetry energy at $\rho_0$~\cite{Chen:2010qx, Kortelainen2010, Chen:2011ps}. Consequently, the Skyrme EDF has the following 10 parameters:
\begin{eqnarray}
\bm{\theta}_{\rm{Sky}} & =& \lbrace  \rho_0,~E_0(\rho_0),~K_0,~E_{\mathrm{sym}}(\rho_0),~L, \notag \\
& &~ G_S,~G_V,~m_{s,0}^{\ast},~m_{v,0}^{\ast},~W_0\rbrace.
\end{eqnarray}
 {For given $\bm{\theta}_{\rm{Sky}} $, the ground-state properties of even-even nuclei are calculated using the Skryme-Hartree-Fock code reported in \cite{Reinhard1991} with the pairing correlation included via the constant-gap approach. }

\begin{figure*}[hbtp]
  \includegraphics[width=1\textwidth]{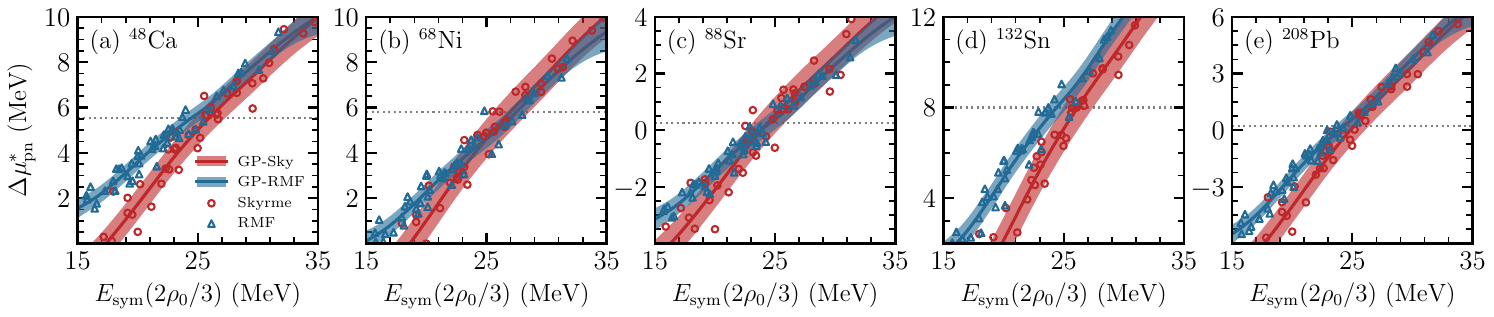}
  \caption{Effective proton-neutron chemical potential differences $\Delta \mu^*_{\rm{pn}}$  of $^{48}\text{Ca}$ (a), $^{68}\text{Ni}$ (b), $^{88}\text{Sr}$ (c), $^{132}\text{Sn}$ (d) and $^{208}\text{Pb}$ (e) versus  $E_{\mathrm{sym}}(2\rho_0/3)$  predicted by the sampled Skyrme EDFs (open cycles) and nonlinear RMF model parameter sets (open triangles). The predictions of Gaussian Process emulators trained using the Skyrme and RMF parameter sets are respectively shown as the {red} and {blue} lines. The shaded bands present the $1\sigma$ error bars of GP predictions. The experimental values~\cite{AME2020} for the $\Delta \mu^*_{\rm{pn}}$ are indicated by the grey dotted lines. \label{fig:GP}}
  \end{figure*}

As for the covariant EDF, we take the nonlinear RMF model based on the effective Lagrangian density\cite{Horowitz:2000xj}:
\begin{equation}\label{lag}
\begin{aligned}
\mathcal{L} = & \bar{\psi}\left(i \partial_\mu \gamma^\mu- m\right) \psi -e \bar{\psi} \gamma_\mu \frac{1+\tau_3}{2} A^\mu \psi-\frac{1}{4} F^{\mu\nu} F_{\mu\nu}\\
& +g_\sigma \sigma \bar{\psi} \psi-g_\omega \omega_\mu \bar{\psi} \gamma^\mu \psi-g_\rho \vec{\rho}_\mu \bar{\psi} \gamma^\mu \vec{\tau} \psi \\
& +\frac{1}{2}\partial_\mu \sigma \partial^\mu \sigma-\frac{1}{2} m_\sigma^2 \sigma^2-\frac{1}{3} b_\sigma M\left(g_\sigma \sigma\right)^3-\frac{1}{4} c_\sigma\left(g_\sigma \sigma\right)^4 \\
& -\frac{1}{4} \omega_{\mu \nu} \omega^{\mu \nu}+\frac{1}{2} m_\omega^2 \omega_\mu \omega^\mu+\frac{1}{4} c_\omega\left(g_\omega^2 \omega_\mu \omega^\mu\right)^2\\
& -\frac{1}{4} \rho_{\mu \nu} \rho^{\mu \nu}+\frac{1}{2} m_\rho^2 \vec{\rho}_\mu \vec{\rho}^\mu+\frac{1}{2} \Lambda_{\rm{V}}\left(g_\rho^2 \vec{\rho}_\mu \vec{\rho}^\mu\right)\left(g_\omega^2 \omega_\mu \omega^\mu\right),
\end{aligned}
\end{equation}
where $\psi$ is the nucleon field, $A^{\mu}$ is the photon field, and $\sigma$, $\omega_{\mu}$, $\vec{\rho}_{\mu}$ represent  the isoscalar-scalar, 
isoscalar-vector and isovector-vector meson fields, respectively.  $\omega_{\mu\nu}=\partial_\mu \omega_\nu-\partial_\nu\omega_\mu$, $\vec{\rho}_{\mu \nu}=\partial_\mu \vec{\rho}_\nu-\partial_\nu \vec{\rho}_\mu$, $F_{\mu \nu}=\partial_\mu A_\nu-\partial_\nu A_\mu$ are the field tensors. 
$\vec{\tau}$ is the isospin Pauli matrices, and we take its third component $\tau_3=1(-1)$ for proton(neutron). 
In addition, $m=939\,\mathrm{MeV}$ is the nucleon bare mass, while $m_{\sigma}$, $m_{\omega}$, $m_{\rho}$ are meson masses. The coupling constants of mesons to nucleons are denoted by $g_{\sigma}$, $g_{\omega}$ and $g_{\rho}$, while $b_{\sigma}$, $c_{\sigma}$ and $c_{\omega}$ describe meson self-coupling  and $\Lambda_{\rm{V}}$ is the $\rho$-$\omega$ coupling constant. In this work, the $m_\omega$ and $m_{\rho}$ are taken to be their experimental values of $m_\omega =782.5 ~\mathrm{MeV}$ and $m_{\rho}= 763  ~\mathrm{MeV}$, respectively.
The adopted RMF model then has 8 adjustable parameters of $m_{\sigma}$, $g_{\sigma}$, $g_{\omega}$, $g_{\rho}$, $b_{\sigma}$, $c_{\sigma}$, $c_{\omega}$ and $\Lambda_{\rm{V}}$. As in the Skyrme EDF, these parameters can be analytically deduced from pseudo-observables in nuclear matter (see., e.g., \cite{Cai:2014kya, Chen:2014}), and we use 
\begin{eqnarray}
  \bm{\theta}_{\rm{RMF}} & =& \lbrace  \rho_0,~E_0(\rho_0),~K_0,~E_{\mathrm{sym}}(\rho_0),~L, \notag \\ 
  & &m_{\rm{Dirac}}^*, m_{\sigma}, c_{\omega}\rbrace,
  \end{eqnarray}
  following Ref.~\cite{Chen:2014}. Here, the $m_{\rm{Dirac}}^*$ is the nucleon Dirac effective mass in symmetric nuclear matter at saturation density. {Based on the Lagrangian density given in Eq.(\ref{lag}), the DIRHF package~\cite{Niksic:2014dra} is modified to calculate the ground-state properties of finite nuclei  by solving the stationary relativistic Hartree-Bogoliubov equations. The pairing force is deduced from the D1S Gogny force (see Ref.~\cite{Niksic:2014dra} for details). }

By assuming the model parameters are uniformly distributed in their empirical ranges~\cite{Chen:2007,Chen:2009wv,Dutra:2012mb,Dutra:2014qga,Sun:2023xkg}, we generate 50 parameter sets for each model from the Latin hypercube sampling. For the common parameters in $\bm{\theta}_{\rm{Sky}}$ and $\bm{\theta}_{\rm{RMF}}$, we take their ranges of variations to be $\rho_0 = 0.155 \pm 0.01~\rm{fm}^{-3}$, $E_0(\rho_0) = -16 \pm 0.6~\rm{MeV}$, $K_0 = 240\pm 20~\rm{MeV}$, $E_{\rm{sym}}(\rho_0)= 34 \pm 6~\rm{MeV}$ and $L=100\pm100~\rm{MeV}$. Concerning the other parameters in $\bm{\theta}_{\rm{Sky}}$ and $\bm{\theta}_{\rm{RMF}}$, their ranges are 
$110\le G_S \le 170~(\rm{MeV}\cdot \rm{fm}^5)$; $-70\le G_V \le 70~(\rm{MeV}\cdot \rm{fm}^5)$; $90\le W_0\le 140~(\rm{MeV}\cdot\rm{fm}^5)$; $0.7\le m_{s,0}^{\ast}/m \le 1$; $0.6\le m_{v,0}^{\ast}/m \le0.9$; $0.55\le m_{\rm{Dirac}}^*/m\le0.65$; $480 \le m_{\sigma} \le 520~(\rm{MeV})$; $0\le c_{\omega}\le 0.012$. Widening these variation ranges by $10\%$  has 
negligible effects on the extracted symmetry energy at subsaturation densities.

Using the sampled 50 Skyrme EDFs and 50 covariant EDFs, we calculate the $\Delta\mu^*_{\rm{pn}}$ of 5 doubly magic nuclei $^{48}\text{Ca}$, $^{68}\text{Ni}$, $^{88}\text{Sr}$, $^{132}\text{Sn}$ and $^{208}\text{Pb}$ as well as the symmetry energy at subsaturation densities. The Pearson correlation coefficients $R$ between 
the $\Delta\mu^*_{\rm{pn}}$ and the $E_{\rm{sym}}(\rho)$ within the Skyrme EDF and the nonlinear RMF model 
are calculated and shown as functions of nucleon density $\rho$ in Fig.~\ref{fig:CorCoeff} (a) and (b), respectively. As can be seen, although the Pearson correlation coefficient depends on specific model and nucleus, they all reach a maximum larger than 0.96 at subsaturation densities of $1/2\le \rho/\rho_0\le3/4$, which indicates a strong linear correlation between the $\Delta\mu^* _{\rm{pn}}$ and the symmetry energy at subsaturation densities.

To see more clearly the $\Delta\mu^*_{\rm{pn}}$-$E_{\rm{sym}}(2\rho_0/3)$ correlations, we 
plot in Fig.~\ref{fig:GP} data-to-data relations predicted by the sampled parameter sets for the Skyrme EDF and the nonlinear RMF model by open circles and triangles, respectively. The experimental values~\cite{AME2020} for the $\Delta\mu^*_{\rm{pn}}$ are indicated by the grey dotted lines. One sees that, the predicted values of $\Delta\mu^*_{\rm{pn}}$
for the chosen 5 doubly magic nuclei are rather sensitive to the $E_{\rm{sym}}(2\rho_0/3)$ within both the two models, although there exist model discrepancies on the predicted $\Delta\mu^*_{\rm{pn}}$-$E_{\rm{sym}}(2\rho_0/3)$ relations.

Given the strong $\Delta\mu^*_{\rm{pn}}$-$E_{\rm{sym}}(2\rho_0/3)$ correlations, we take the predictions 
of the sampled EDFs as training points to construct emulators using the \textit{surmise} python package~\cite{surmise2023} for the $\Delta \mu^*_{\rm{pn}}(N,Z)$ as functions of $E_{\rm{sym}}(2\rho_0/3)$ via Gaussian processes (see, e.g.,~\cite{Bernhard:2015hxa}). Particularly, the effects of the other model parameters  except $E_{\rm{sym}}(2\rho_0/3)$ are mimicked by noise terms in GPs. The GP predictions  from the Skyrme EDF and the nonlinear RMF model  are shown as the red and blue lines labeled by GP-Sky and GP-RMF in Fig.~\ref{fig:GP}, respectively. The shaded region represent the  $1\sigma$ uncertainty bands of GP predictions.
It is seen that the GPs well reproduce the dependence of the $\Delta\mu^*_{\rm{pn}}$  on the
$E_{\rm{sym}}(2\rho_0/3)$  and reasonably describe the model uncertainties.

\begin{figure}[htbp]
  \includegraphics[width=0.5\textwidth]{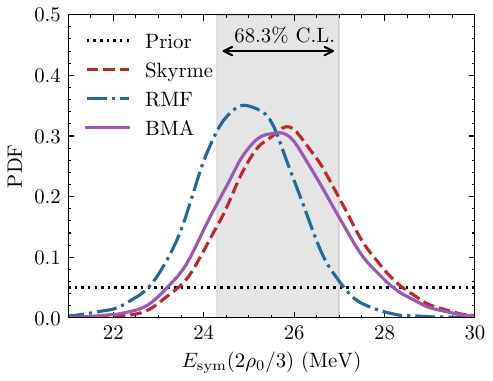}
  \caption{Posterior probability density distribution (PDF) of $E_{\text{sym}}(2\rho_0/3)$ inferred from the Skyrme EDF (red dashed line), the RMF model (blue dashed-dotted line), and the Bayesian  model averaging (purple solid line). 
  The prior distribution of $E_{\text{sym}}(2\rho_0/3)$ 
  is plotted as the black dotted line.  The grey shaded area indicates the 68.3$\%$ credible interval of posterior from BMA.\label{fig:post}}
  \end{figure}

Based on the GPs constructed from the Skyrme EDFs and the nonlinear RMF models, we further extract the $E_{\mathrm{sym}}(2\rho_0/3)$ from Bayesian analyses of  the $\Delta \mu^*_{\rm{pn}}$ of 5 doubly magic nuclei,  $^{48}\text{Ca},^{68}\text{Ni},^{88}\text{Sr}$, $^{132}\text{Sn}$ and $^{208}\text{Pb}$. The prior of $E_{\mathrm{sym}}(2\rho_0/3)$  is taken 
to be uniform in the range of $15$-$35$ MeV, and the likelihood is given by 
$$p(\bm{y}\vert \theta, \sigma)\propto \exp \left[-\frac{1}{2}\left(\boldsymbol{y}-\boldsymbol{y}^{\rm{exp}}\right)^T \Sigma^{-1}\left(\boldsymbol{y}-\boldsymbol{y}^{\rm{exp}}\right)\right],
$$
where the model parameter $\theta$ is $E_{\rm{sym}}(2\rho_0/3)$, and $\bm{y}$ and $\bm{y}^{\rm{exp}}$ are the GP predictions and experimental values for the $\Delta \mu^*_{\rm{pn}}$ of the 5 doubly magic nuclei, respectively. The covariance matrix ${\Sigma} = {\Sigma}_{\rm{GP}} + \rm{diag}(\sigma^2)$ includes the uncertainties and correlations of the emulator predictions via the GP covariance matrix ${\Sigma}_{\rm{GP}} $, and the adopted error $\sigma$ which takes into account the deficiency of theoretical models and experimental errors. In prior, the model deficiency is unknown and we assume that $\sigma$ lies uniformly between 0 and 10 MeV. The posterior distributions of $E_{\rm{sym}}(2\rho_0/3)$ and $\sigma$ are then generated  using sequential Monte Carlo algorithm from PyMCv4.0\cite{abril2023pymc}. Based on GP-Sky and GP-RMF, the posterior $\sigma$ are inferred to be 
$0.4_{-0.2}^{+0.4}$ and $0.8_{-0.3}^{+0.6}$ MeV, respectively.
The smaller $\sigma$ value of GP-Sky indicates the Skyrme EDF is more capable of simultaneously reproducing the $\Delta\mu^*_{\rm{pn}}$ of the 5 doubly magic nuclei. The posterior distributions of $E_{\rm{sym}}(2\rho_0/3)$ from GP-Sky and GP-RMF are shown  in Fig.~\ref{fig:post} as the red dashed and blue dashed-dotted lines, respectively. Quantitatively, we obtain  $E_{\rm{sym}}(2\rho_0/3)=25.8_{-1.2}^{+1.3}$ MeV and $E_{\rm{sym}}(2\rho_0/3)=24.9\pm 1.1$ MeV at $68.3\%$ confidence level from the Skyrme and RMF models, respectively.

To handle with the model dependence of the inferred $E_{\rm{sym}}(2\rho_0/3)$, we further carried out  Bayesian model averaging\cite{RevModPhys.83.943,Kejzlar_2020,Jay:2020jkz} over the results from the two models. Within the BMA approach, the posterior distribution of any quantity 
$\mathcal{O}$ is obtained by mixing the posterior predictive distributions $p(\mathcal{O}\vert \bm{y}, \mathcal{M}_i)$ of each individual model $\mathcal{M}_i$ with a probability weight $p(\mathcal{M}_i\vert \bm{y})$, i.e.,
\begin{eqnarray}\label{eq:bma}
  p\left(\mathcal{O}\vert \bm{y}\right)=\sum_i p(\mathcal{O}\vert \bm{y}, \mathcal{M}_i)p(\mathcal{M}_i\vert \bm{y}).
\end{eqnarray}
The model weight, reflecting the validity of $\mathcal{M}_i$ given experimental data $\bm{y}$,  is given by
\begin{equation}\label{eq:wt}
  \begin{aligned}
  p\left(\mathcal{M}_i \mid \boldsymbol{y}\right)&=\frac{p\left(\boldsymbol{y} \mid \mathcal{M}_i\right) \pi\left(\mathcal{M}_i\right)}{\sum_{\ell} p\left(\boldsymbol{y} \mid \mathcal{M}_{\ell}\right) \pi\left(\mathcal{M}_{\ell}\right)},
  \end{aligned}
  \end{equation}
where the prior model probability $\pi (\mathcal{M}_i)$ reflects our preference on $\mathcal{M}_i$ before seeing the data, and the evidence $p\left(\boldsymbol{y} \mid \mathcal{M}_i\right)$ measures the probability that the model reproduces the experimental data $\bm{y}$ with given prior distributions of model parameter and adopted errors $\pi(\theta, \sigma\vert \mathcal{M}_i)$, i.e., 
\begin{equation}
  p\left(\boldsymbol{y} \mid \mathcal{M}_i\right) = \int p(\bm{y}\vert \theta_i, \sigma_i,\mathcal{M}_i) \pi(\theta, \sigma\vert \mathcal{M}_i)d\theta_i d\sigma_i.
\end{equation} 
In our case, in prior, the Skryme and nonlinear RMF models are equally weighted with
$\pi (\mathcal{M}_i) = 1/2$, and 
the evidences $p\left(\boldsymbol{y} \mid \mathcal{M}_i\right)$ of the two models are estimated with the sequential Monte Carlo method. The obtained ratio of the GP-Sky evidence to that of GP-RMF is estimated to be 3.3, which means the measured $\Delta\mu^*_{\rm{pn}}$ for the chosen 5 doubly magic nuclei slightly prefers the Skyrme EDF rather than the nonlinear RMF model. Invoking Eqs.(\ref{eq:bma}) and (\ref{eq:wt}), we calculate the BMA posterior distribution of the $E_{\rm{sym}}(2\rho_0/3)$ based on those from the Skyrme EDF and nonlinear RMF model, and show it as the purple solid line in Fig.~\ref{fig:post}. Quantitatively,  the $E_{\rm{sym}}(2\rho_0/3)$ is inferred to be {$25.6_{-1.3}^{+1.4}$} MeV at 1$\sigma$ confidence level from the BMA analysis.

\begin{figure}[htbp]
  \includegraphics[width=0.5\textwidth]{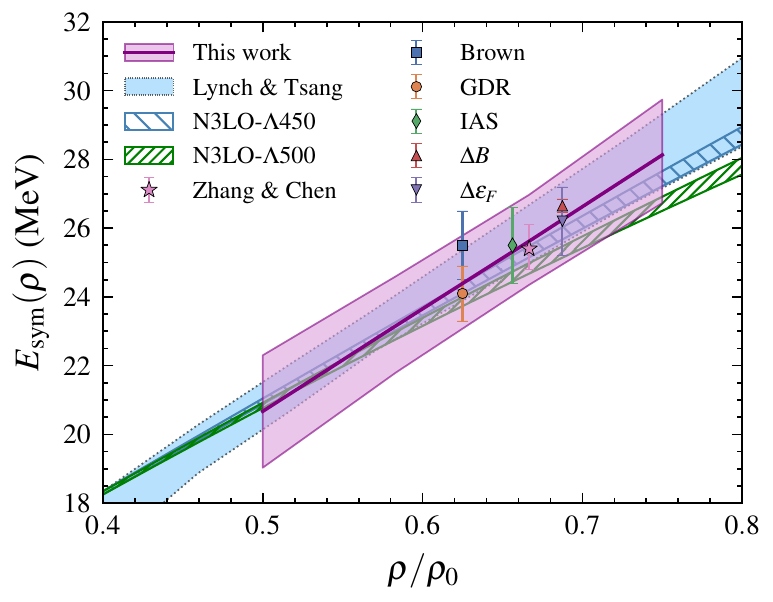}
  \caption{Constraints on the symmetry energy as a function of density $\rho$ from Bayesian model averaging of Skyrme and RMF model predictions. Various experimental constraints and theoretical predictions are shown for comparison (see text for details). \label{fig:Erho}}
  \end{figure}

From similar analyses, we extract the symmetry energy from $\rho_0/2$ to {$3\rho_0/4$} by BMA over the predictions of the Skyrme EDF and the nonlinear RMF model, and show in Fig.~\ref{fig:Erho} the median value and the 1$\sigma$ confidence region as the purple solid line and shaded band, respectively. For comparison,
Fig.~\ref{fig:Erho} also includes the results from fitting various terrestrial and astrophysical constraints with a cubic polynomical form of the potential part of symmetry energy (Lynch $\&$ Tsang)~\cite{Lynch:2021xkq}, and the predictions of many-body perturbation calculations using 
N3LO chiral interactions with momentum cutoff $\Lambda=450$ and $500$ MeV (N3LO-$\Lambda450$ and N3LO-$\Lambda500$ )~\cite{Drischler:2020hwi}. Note that for the N3LO-$\Lambda450$ and N3LO-$\Lambda500$ results, the saturation density is taken to be their respective predicted values of $0.17$ and $0.173~\rm{fm}^{-3}$. In addition, we also exhibit as symbols in Fig.~\ref{fig:Erho} 
 six constraints on the symmetry energy around $2\rho_0/3$ extracted from the properties of doubly magic nuclei (Brown)~\cite{Brown:2013mga}, and the giant dipole resonance (GDR) in $^{208}$Pb~\cite{Trippa:2008gr}, the binding energy difference of heavy isotope pairs ($\Delta B$)~\cite{Zhang:2013wna}, the neutron-proton Fermi energy difference ($\Delta \epsilon_F$)~\cite{Wang:2013}, and a recent Skyrme EDF analysis of PREX-2 and CREX results on the charge-weak form factor difference in $^{208}$Pb and $^{48}$Ca together with properties of doubly magic nuclei (Zhang $\&$ Chen)~
\cite{Zhang:2022bni}. Our results agree well with these constraints and predictions, and provide more statistically reliable predictions.

In conclusion, based on a number of Skyrme energy density functionals and nonlinear relativistic mean field model parameter sets, we extract the symmetry energy $E_{\mathrm{sym}}(\rho)$ around $2\rho_0/3$ from a Bayesian model averaging analysis of 
the effective proton-neutron chemical potential difference $\Delta\mu^*_{\rm{pn}}$ for  5
doubly magic nuclei, $^{48}$Ca, $^{68}$Ni, $^{88}$Sr, $^{132}$Sn and $^{208}$Pb. Particularly,  we infer $E_{\mathrm{sym}}(2\rho_0/3) =  25.6_{-1.3}^{+1.4}~\mathrm{MeV}$ at $1\sigma$ confidence level.
The obtained constraints on the $E_{\mathrm{sym}}(\rho)$ around $2\rho_0/3$ are consistent with microscopic predictions and results from analyses of other isovector indicators, and  are potentially useful in the studies of, e.g., the neutron drip line, r-process path, the inner crust of neutron stars and the core-collapse supernovae\cite{Wang:2014mra,Oertel:2016bki}.

Since both the intra- and inter-model uncertainties are taken into account in our BMA analyses, the present results are statistically more reliable. Moreover, allowing further inclusion of more experimental observables and theoretical models, the employed BMA method provides a rigorous statistical approach towards the accurate determination of nuclear symmetry energy.

\begin{acknowledgments}
 M. Qiu thanks Meng-Lan Liu and Bo-Shuai Cai for helpful discussions.   This work was supported in part by the National Natural Science Foundation of China under Grant Nos. 12235010, 11905302 and 11625521,  the National SKA Program of China No. 2020SKA0120300, and the Guangdong Major Project of Basic and Applied Basic Research under Grant No. 2021B0301030006.

\end{acknowledgments}

\bibliography{ref}

\providecommand{\noopsort}[1]{}\providecommand{\singleletter}[1]{#1}%
\begin{thebibliography}{66}%
\makeatletter
\providecommand \@ifxundefined [1]{%
 \@ifx{#1\undefined}
}%
\providecommand \@ifnum [1]{%
 \ifnum #1\expandafter \@firstoftwo
 \else \expandafter \@secondoftwo
 \fi
}%
\providecommand \@ifx [1]{%
 \ifx #1\expandafter \@firstoftwo
 \else \expandafter \@secondoftwo
 \fi
}%
\providecommand \natexlab [1]{#1}%
\providecommand \enquote  [1]{``#1''}%
\providecommand \bibnamefont  [1]{#1}%
\providecommand \bibfnamefont [1]{#1}%
\providecommand \citenamefont [1]{#1}%
\providecommand \href@noop [0]{\@secondoftwo}%
\providecommand \href [0]{\begingroup \@sanitize@url \@href}%
\providecommand \@href[1]{\@@startlink{#1}\@@href}%
\providecommand \@@href[1]{\endgroup#1\@@endlink}%
\providecommand \@sanitize@url [0]{\catcode `\\12\catcode `\$12\catcode
  `\&12\catcode `\#12\catcode `\^12\catcode `\_12\catcode `\%12\relax}%
\providecommand \@@startlink[1]{}%
\providecommand \@@endlink[0]{}%
\providecommand \url  [0]{\begingroup\@sanitize@url \@url }%
\providecommand \@url [1]{\endgroup\@href {#1}{\urlprefix }}%
\providecommand \urlprefix  [0]{URL }%
\providecommand \Eprint [0]{\href }%
\providecommand \doibase [0]{https://doi.org/}%
\providecommand \selectlanguage [0]{\@gobble}%
\providecommand \bibinfo  [0]{\@secondoftwo}%
\providecommand \bibfield  [0]{\@secondoftwo}%
\providecommand \translation [1]{[#1]}%
\providecommand \BibitemOpen [0]{}%
\providecommand \bibitemStop [0]{}%
\providecommand \bibitemNoStop [0]{.\EOS\space}%
\providecommand \EOS [0]{\spacefactor3000\relax}%
\providecommand \BibitemShut  [1]{\csname bibitem#1\endcsname}%
\let\auto@bib@innerbib\@empty
\bibitem [{\citenamefont {Baran}\ \emph {et~al.}(2005)\citenamefont {Baran},
  \citenamefont {Colonna}, \citenamefont {Greco},\ and\ \citenamefont {{Di
  Toro}}}]{Baran2005}%
  \BibitemOpen
  \bibfield  {author} {\bibinfo {author} {\bibfnamefont {V.}~\bibnamefont
  {Baran}}, \bibinfo {author} {\bibfnamefont {M.}~\bibnamefont {Colonna}},
  \bibinfo {author} {\bibfnamefont {V.}~\bibnamefont {Greco}},\ and\ \bibinfo
  {author} {\bibfnamefont {M.}~\bibnamefont {{Di Toro}}},\ }\href
  {https://doi.org/10.1016/j.physrep.2004.12.004} {\bibfield  {journal}
  {\bibinfo  {journal} {Phys. Rep.}\ }\textbf {\bibinfo {volume} {410}},\
  \bibinfo {pages} {335} (\bibinfo {year} {2005})}\BibitemShut {NoStop}%
\bibitem [{\citenamefont {Steiner}\ \emph {et~al.}(2005)\citenamefont
  {Steiner}, \citenamefont {Prakash}, \citenamefont {Lattimer},\ and\
  \citenamefont {Ellis}}]{Steiner2005b}%
  \BibitemOpen
  \bibfield  {author} {\bibinfo {author} {\bibfnamefont {A.~W.}\ \bibnamefont
  {Steiner}}, \bibinfo {author} {\bibfnamefont {M.}~\bibnamefont {Prakash}},
  \bibinfo {author} {\bibfnamefont {J.~M.}\ \bibnamefont {Lattimer}},\ and\
  \bibinfo {author} {\bibfnamefont {P.~J.}\ \bibnamefont {Ellis}},\ }\href
  {https://doi.org/10.1016/j.physrep.2005.02.004} {\bibfield  {journal}
  {\bibinfo  {journal} {Phys. Rep.}\ }\textbf {\bibinfo {volume} {411}},\
  \bibinfo {pages} {325} (\bibinfo {year} {2005})}\BibitemShut {NoStop}%
\bibitem [{\citenamefont {Lattimer}\ and\ \citenamefont
  {Prakash}(2007)}]{Lattimer:2006xb}%
  \BibitemOpen
  \bibfield  {author} {\bibinfo {author} {\bibfnamefont {J.~M.}\ \bibnamefont
  {Lattimer}}\ and\ \bibinfo {author} {\bibfnamefont {M.}~\bibnamefont
  {Prakash}},\ }\href {https://doi.org/10.1016/j.physrep.2007.02.003}
  {\bibfield  {journal} {\bibinfo  {journal} {Phys. Rept.}\ }\textbf {\bibinfo
  {volume} {442}},\ \bibinfo {pages} {109} (\bibinfo {year}
  {2007})}\BibitemShut {NoStop}%
\bibitem [{\citenamefont {Li}\ \emph {et~al.}(2008)\citenamefont {Li},
  \citenamefont {Chen},\ and\ \citenamefont {Ko}}]{Li:2008gp}%
  \BibitemOpen
  \bibfield  {author} {\bibinfo {author} {\bibfnamefont {B.-A.}\ \bibnamefont
  {Li}}, \bibinfo {author} {\bibfnamefont {L.-W.}\ \bibnamefont {Chen}},\ and\
  \bibinfo {author} {\bibfnamefont {C.~M.}\ \bibnamefont {Ko}},\ }\href
  {https://doi.org/10.1016/j.physrep.2008.04.005} {\bibfield  {journal}
  {\bibinfo  {journal} {Phys. Rept.}\ }\textbf {\bibinfo {volume} {464}},\
  \bibinfo {pages} {113} (\bibinfo {year} {2008})}\BibitemShut {NoStop}%
\bibitem [{\citenamefont {Baldo}\ and\ \citenamefont
  {Burgio}(2016)}]{Baldo:2016jhp}%
  \BibitemOpen
  \bibfield  {author} {\bibinfo {author} {\bibfnamefont {M.}~\bibnamefont
  {Baldo}}\ and\ \bibinfo {author} {\bibfnamefont {G.~F.}\ \bibnamefont
  {Burgio}},\ }\href {https://doi.org/10.1016/j.ppnp.2016.06.006} {\bibfield
  {journal} {\bibinfo  {journal} {Prog. Part. Nucl. Phys.}\ }\textbf {\bibinfo
  {volume} {91}},\ \bibinfo {pages} {203} (\bibinfo {year} {2016})}\BibitemShut
  {NoStop}%
\bibitem [{\citenamefont {Oertel}\ \emph {et~al.}(2017)\citenamefont {Oertel},
  \citenamefont {Hempel}, \citenamefont {Kl\"ahn},\ and\ \citenamefont
  {Typel}}]{Oertel:2016bki}%
  \BibitemOpen
  \bibfield  {author} {\bibinfo {author} {\bibfnamefont {M.}~\bibnamefont
  {Oertel}}, \bibinfo {author} {\bibfnamefont {M.}~\bibnamefont {Hempel}},
  \bibinfo {author} {\bibfnamefont {T.}~\bibnamefont {Kl\"ahn}},\ and\ \bibinfo
  {author} {\bibfnamefont {S.}~\bibnamefont {Typel}},\ }\href
  {https://doi.org/10.1103/RevModPhys.89.015007} {\bibfield  {journal}
  {\bibinfo  {journal} {Rev. Mod. Phys.}\ }\textbf {\bibinfo {volume} {89}},\
  \bibinfo {pages} {015007} (\bibinfo {year} {2017})}\BibitemShut {NoStop}%
\bibitem [{\citenamefont {Roca-Maza}\ and\ \citenamefont
  {Paar}(2018)}]{Roca-Maza:2018ujj}%
  \BibitemOpen
  \bibfield  {author} {\bibinfo {author} {\bibfnamefont {X.}~\bibnamefont
  {Roca-Maza}}\ and\ \bibinfo {author} {\bibfnamefont {N.}~\bibnamefont
  {Paar}},\ }\href {https://doi.org/10.1016/j.ppnp.2018.04.001} {\bibfield
  {journal} {\bibinfo  {journal} {Prog. Part. Nucl. Phys.}\ }\textbf {\bibinfo
  {volume} {101}},\ \bibinfo {pages} {96} (\bibinfo {year} {2018})}\BibitemShut
  {NoStop}%
\bibitem [{\citenamefont {Li}\ \emph {et~al.}(2021)\citenamefont {Li},
  \citenamefont {Cai}, \citenamefont {Xie},\ and\ \citenamefont
  {Zhang}}]{Li:2021thg}%
  \BibitemOpen
  \bibfield  {author} {\bibinfo {author} {\bibfnamefont {B.-A.}\ \bibnamefont
  {Li}}, \bibinfo {author} {\bibfnamefont {B.-J.}\ \bibnamefont {Cai}},
  \bibinfo {author} {\bibfnamefont {W.-J.}\ \bibnamefont {Xie}},\ and\ \bibinfo
  {author} {\bibfnamefont {N.-B.}\ \bibnamefont {Zhang}},\ }\href
  {https://doi.org/10.3390/universe7060182} {\bibfield  {journal} {\bibinfo
  {journal} {Universe}\ }\textbf {\bibinfo {volume} {7}},\ \bibinfo {pages}
  {182} (\bibinfo {year} {2021})}\BibitemShut {NoStop}%
\bibitem [{\citenamefont {Adhikari}\ \emph {et~al.}(2021)\citenamefont
  {Adhikari} \emph {et~al.}}]{PREX:2021umo}%
  \BibitemOpen
  \bibfield  {author} {\bibinfo {author} {\bibfnamefont {D.}~\bibnamefont
  {Adhikari}} \emph {et~al.} (\bibinfo {collaboration} {PREX}),\ }\href
  {https://doi.org/10.1103/PhysRevLett.126.172502} {\bibfield  {journal}
  {\bibinfo  {journal} {Phys. Rev. Lett.}\ }\textbf {\bibinfo {volume} {126}},\
  \bibinfo {pages} {172502} (\bibinfo {year} {2021})}\BibitemShut {NoStop}%
\bibitem [{\citenamefont {Adhikari}\ \emph {et~al.}(2022)\citenamefont
  {Adhikari} \emph {et~al.}}]{CREX:2022kgg}%
  \BibitemOpen
  \bibfield  {author} {\bibinfo {author} {\bibfnamefont {D.}~\bibnamefont
  {Adhikari}} \emph {et~al.} (\bibinfo {collaboration} {CREX}),\ }\href
  {https://doi.org/10.1103/PhysRevLett.129.042501} {\bibfield  {journal}
  {\bibinfo  {journal} {Phys. Rev. Lett.}\ }\textbf {\bibinfo {volume} {129}},\
  \bibinfo {pages} {042501} (\bibinfo {year} {2022})}\BibitemShut {NoStop}%
\bibitem [{\citenamefont {Reed}\ \emph {et~al.}(2021)\citenamefont {Reed},
  \citenamefont {Fattoyev}, \citenamefont {Horowitz},\ and\ \citenamefont
  {Piekarewicz}}]{Reed:2021nqk}%
  \BibitemOpen
  \bibfield  {author} {\bibinfo {author} {\bibfnamefont {B.~T.}\ \bibnamefont
  {Reed}}, \bibinfo {author} {\bibfnamefont {F.~J.}\ \bibnamefont {Fattoyev}},
  \bibinfo {author} {\bibfnamefont {C.~J.}\ \bibnamefont {Horowitz}},\ and\
  \bibinfo {author} {\bibfnamefont {J.}~\bibnamefont {Piekarewicz}},\ }\href
  {https://doi.org/10.1103/PhysRevLett.126.172503} {\bibfield  {journal}
  {\bibinfo  {journal} {Phys. Rev. Lett.}\ }\textbf {\bibinfo {volume} {126}},\
  \bibinfo {pages} {172503} (\bibinfo {year} {2021})}\BibitemShut {NoStop}%
\bibitem [{\citenamefont {Piekarewicz}(2021)}]{Piekarewicz:2021jte}%
  \BibitemOpen
  \bibfield  {author} {\bibinfo {author} {\bibfnamefont {J.}~\bibnamefont
  {Piekarewicz}},\ }\href {https://doi.org/10.1103/PhysRevC.104.024329}
  {\bibfield  {journal} {\bibinfo  {journal} {Phys. Rev. C}\ }\textbf {\bibinfo
  {volume} {104}},\ \bibinfo {pages} {024329} (\bibinfo {year}
  {2021})}\BibitemShut {NoStop}%
\bibitem [{\citenamefont {Reinhard}\ \emph {et~al.}(2021)\citenamefont
  {Reinhard}, \citenamefont {Roca-Maza},\ and\ \citenamefont
  {Nazarewicz}}]{Reinhard:2021utv}%
  \BibitemOpen
  \bibfield  {author} {\bibinfo {author} {\bibfnamefont {P.-G.}\ \bibnamefont
  {Reinhard}}, \bibinfo {author} {\bibfnamefont {X.}~\bibnamefont
  {Roca-Maza}},\ and\ \bibinfo {author} {\bibfnamefont {W.}~\bibnamefont
  {Nazarewicz}},\ }\href {https://doi.org/10.1103/PhysRevLett.127.232501}
  {\bibfield  {journal} {\bibinfo  {journal} {Phys. Rev. Lett.}\ }\textbf
  {\bibinfo {volume} {127}},\ \bibinfo {pages} {232501} (\bibinfo {year}
  {2021})}\BibitemShut {NoStop}%
\bibitem [{\citenamefont {Reinhard}\ \emph {et~al.}(2022)\citenamefont
  {Reinhard}, \citenamefont {Roca-Maza},\ and\ \citenamefont
  {Nazarewicz}}]{Reinhard:2022inh}%
  \BibitemOpen
  \bibfield  {author} {\bibinfo {author} {\bibfnamefont {P.-G.}\ \bibnamefont
  {Reinhard}}, \bibinfo {author} {\bibfnamefont {X.}~\bibnamefont
  {Roca-Maza}},\ and\ \bibinfo {author} {\bibfnamefont {W.}~\bibnamefont
  {Nazarewicz}},\ }\href {https://doi.org/10.1103/PhysRevLett.129.232501}
  {\bibfield  {journal} {\bibinfo  {journal} {Phys. Rev. Lett.}\ }\textbf
  {\bibinfo {volume} {129}},\ \bibinfo {pages} {232501} (\bibinfo {year}
  {2022})}\BibitemShut {NoStop}%
\bibitem [{\citenamefont {Y\"uksel}\ and\ \citenamefont
  {Paar}(2023)}]{Yuksel:2022umn}%
  \BibitemOpen
  \bibfield  {author} {\bibinfo {author} {\bibfnamefont {E.}~\bibnamefont
  {Y\"uksel}}\ and\ \bibinfo {author} {\bibfnamefont {N.}~\bibnamefont
  {Paar}},\ }\href {https://doi.org/10.1016/j.physletb.2022.137622} {\bibfield
  {journal} {\bibinfo  {journal} {Phys. Lett. B}\ }\textbf {\bibinfo {volume}
  {836}},\ \bibinfo {pages} {137622} (\bibinfo {year} {2023})}\BibitemShut
  {NoStop}%
\bibitem [{\citenamefont {Zhang}\ and\ \citenamefont
  {Chen}(2023)}]{Zhang:2022bni}%
  \BibitemOpen
  \bibfield  {author} {\bibinfo {author} {\bibfnamefont {Z.}~\bibnamefont
  {Zhang}}\ and\ \bibinfo {author} {\bibfnamefont {L.-W.}\ \bibnamefont
  {Chen}},\ }\href {https://doi.org/10.1103/PhysRevC.108.024317} {\bibfield
  {journal} {\bibinfo  {journal} {Phys. Rev. C}\ }\textbf {\bibinfo {volume}
  {108}},\ \bibinfo {pages} {024317} (\bibinfo {year} {2023})}\BibitemShut
  {NoStop}%
\bibitem [{\citenamefont {Brown}(2000)}]{Brown:2000pd}%
  \BibitemOpen
  \bibfield  {author} {\bibinfo {author} {\bibfnamefont {B.~A.}\ \bibnamefont
  {Brown}},\ }\href {https://doi.org/10.1103/PhysRevLett.85.5296} {\bibfield
  {journal} {\bibinfo  {journal} {Phys. Rev. Lett.}\ }\textbf {\bibinfo
  {volume} {85}},\ \bibinfo {pages} {5296} (\bibinfo {year}
  {2000})}\BibitemShut {NoStop}%
\bibitem [{\citenamefont {Typel}\ and\ \citenamefont
  {Brown}(2001)}]{Typel:2001lcw}%
  \BibitemOpen
  \bibfield  {author} {\bibinfo {author} {\bibfnamefont {S.}~\bibnamefont
  {Typel}}\ and\ \bibinfo {author} {\bibfnamefont {B.~A.}\ \bibnamefont
  {Brown}},\ }\href {https://doi.org/10.1103/PhysRevC.64.027302} {\bibfield
  {journal} {\bibinfo  {journal} {Phys. Rev. C}\ }\textbf {\bibinfo {volume}
  {64}},\ \bibinfo {pages} {027302} (\bibinfo {year} {2001})}\BibitemShut
  {NoStop}%
\bibitem [{\citenamefont {Furnstahl}(2002)}]{Furnstahl:2001un}%
  \BibitemOpen
  \bibfield  {author} {\bibinfo {author} {\bibfnamefont {R.~J.}\ \bibnamefont
  {Furnstahl}},\ }\href {https://doi.org/10.1016/S0375-9474(02)00867-9}
  {\bibfield  {journal} {\bibinfo  {journal} {Nucl. Phys. A}\ }\textbf
  {\bibinfo {volume} {706}},\ \bibinfo {pages} {85} (\bibinfo {year}
  {2002})}\BibitemShut {NoStop}%
\bibitem [{\citenamefont {Chen}\ \emph {et~al.}(2005)\citenamefont {Chen},
  \citenamefont {Ko},\ and\ \citenamefont {Li}}]{Chen:2005ti}%
  \BibitemOpen
  \bibfield  {author} {\bibinfo {author} {\bibfnamefont {L.-W.}\ \bibnamefont
  {Chen}}, \bibinfo {author} {\bibfnamefont {C.~M.}\ \bibnamefont {Ko}},\ and\
  \bibinfo {author} {\bibfnamefont {B.-A.}\ \bibnamefont {Li}},\ }\href
  {https://doi.org/10.1103/PhysRevC.72.064309} {\bibfield  {journal} {\bibinfo
  {journal} {Phys. Rev. C}\ }\textbf {\bibinfo {volume} {72}},\ \bibinfo
  {pages} {064309} (\bibinfo {year} {2005})}\BibitemShut {NoStop}%
\bibitem [{\citenamefont {Roca-Maza}\ \emph {et~al.}(2013)\citenamefont
  {Roca-Maza}, \citenamefont {Centelles}, \citenamefont {Vi\~nas},
  \citenamefont {Brenna}, \citenamefont {Col\`o}, \citenamefont {Agrawal},
  \citenamefont {Paar}, \citenamefont {Piekarewicz},\ and\ \citenamefont
  {Vretenar}}]{Roca-Maza:2013mla}%
  \BibitemOpen
  \bibfield  {author} {\bibinfo {author} {\bibfnamefont {X.}~\bibnamefont
  {Roca-Maza}}, \bibinfo {author} {\bibfnamefont {M.}~\bibnamefont
  {Centelles}}, \bibinfo {author} {\bibfnamefont {X.}~\bibnamefont {Vi\~nas}},
  \bibinfo {author} {\bibfnamefont {M.}~\bibnamefont {Brenna}}, \bibinfo
  {author} {\bibfnamefont {G.}~\bibnamefont {Col\`o}}, \bibinfo {author}
  {\bibfnamefont {B.~K.}\ \bibnamefont {Agrawal}}, \bibinfo {author}
  {\bibfnamefont {N.}~\bibnamefont {Paar}}, \bibinfo {author} {\bibfnamefont
  {J.}~\bibnamefont {Piekarewicz}},\ and\ \bibinfo {author} {\bibfnamefont
  {D.}~\bibnamefont {Vretenar}},\ }\href
  {https://doi.org/10.1103/PhysRevC.88.024316} {\bibfield  {journal} {\bibinfo
  {journal} {Phys. Rev. C}\ }\textbf {\bibinfo {volume} {88}},\ \bibinfo
  {pages} {024316} (\bibinfo {year} {2013})}\BibitemShut {NoStop}%
\bibitem [{\citenamefont {Zhang}\ and\ \citenamefont
  {Chen}(2014)}]{Zhang:2014yfa}%
  \BibitemOpen
  \bibfield  {author} {\bibinfo {author} {\bibfnamefont {Z.}~\bibnamefont
  {Zhang}}\ and\ \bibinfo {author} {\bibfnamefont {L.-W.}\ \bibnamefont
  {Chen}},\ }\href {https://doi.org/10.1103/PhysRevC.90.064317} {\bibfield
  {journal} {\bibinfo  {journal} {Phys. Rev. C}\ }\textbf {\bibinfo {volume}
  {90}},\ \bibinfo {pages} {064317} (\bibinfo {year} {2014})}\BibitemShut
  {NoStop}%
\bibitem [{\citenamefont {Zhang}\ and\ \citenamefont
  {Chen}(2015)}]{Zhang:2015ava}%
  \BibitemOpen
  \bibfield  {author} {\bibinfo {author} {\bibfnamefont {Z.}~\bibnamefont
  {Zhang}}\ and\ \bibinfo {author} {\bibfnamefont {L.-W.}\ \bibnamefont
  {Chen}},\ }\href {https://doi.org/10.1103/PhysRevC.92.031301} {\bibfield
  {journal} {\bibinfo  {journal} {Phys. Rev. C}\ }\textbf {\bibinfo {volume}
  {92}},\ \bibinfo {pages} {031301} (\bibinfo {year} {2015})}\BibitemShut
  {NoStop}%
\bibitem [{\citenamefont {Roca-Maza}\ \emph {et~al.}(2015)\citenamefont
  {Roca-Maza}, \citenamefont {Vi\~nas}, \citenamefont {Centelles},
  \citenamefont {Agrawal}, \citenamefont {Col\'o}, \citenamefont {Paar},
  \citenamefont {Piekarewicz},\ and\ \citenamefont
  {Vretenar}}]{Roca-Maza:2015eza}%
  \BibitemOpen
  \bibfield  {author} {\bibinfo {author} {\bibfnamefont {X.}~\bibnamefont
  {Roca-Maza}}, \bibinfo {author} {\bibfnamefont {X.}~\bibnamefont {Vi\~nas}},
  \bibinfo {author} {\bibfnamefont {M.}~\bibnamefont {Centelles}}, \bibinfo
  {author} {\bibfnamefont {B.~K.}\ \bibnamefont {Agrawal}}, \bibinfo {author}
  {\bibfnamefont {G.}~\bibnamefont {Col\'o}}, \bibinfo {author} {\bibfnamefont
  {N.}~\bibnamefont {Paar}}, \bibinfo {author} {\bibfnamefont {J.}~\bibnamefont
  {Piekarewicz}},\ and\ \bibinfo {author} {\bibfnamefont {D.}~\bibnamefont
  {Vretenar}},\ }\href {https://doi.org/10.1103/PhysRevC.92.064304} {\bibfield
  {journal} {\bibinfo  {journal} {Phys. Rev. C}\ }\textbf {\bibinfo {volume}
  {92}},\ \bibinfo {pages} {064304} (\bibinfo {year} {2015})}\BibitemShut
  {NoStop}%
\bibitem [{\citenamefont {Xiao}\ \emph {et~al.}(2009)\citenamefont {Xiao},
  \citenamefont {Li}, \citenamefont {Chen}, \citenamefont {Yong},\ and\
  \citenamefont {Zhang}}]{Xiao:2008vm}%
  \BibitemOpen
  \bibfield  {author} {\bibinfo {author} {\bibfnamefont {Z.}~\bibnamefont
  {Xiao}}, \bibinfo {author} {\bibfnamefont {B.-A.}\ \bibnamefont {Li}},
  \bibinfo {author} {\bibfnamefont {L.-W.}\ \bibnamefont {Chen}}, \bibinfo
  {author} {\bibfnamefont {G.-C.}\ \bibnamefont {Yong}},\ and\ \bibinfo
  {author} {\bibfnamefont {M.}~\bibnamefont {Zhang}},\ }\href
  {https://doi.org/10.1103/PhysRevLett.102.062502} {\bibfield  {journal}
  {\bibinfo  {journal} {Phys. Rev. Lett.}\ }\textbf {\bibinfo {volume} {102}},\
  \bibinfo {pages} {062502} (\bibinfo {year} {2009})}\BibitemShut {NoStop}%
\bibitem [{\citenamefont {Feng}\ and\ \citenamefont {Jin}(2010)}]{Feng:2009am}%
  \BibitemOpen
  \bibfield  {author} {\bibinfo {author} {\bibfnamefont {Z.-Q.}\ \bibnamefont
  {Feng}}\ and\ \bibinfo {author} {\bibfnamefont {G.-M.}\ \bibnamefont {Jin}},\
  }\href {https://doi.org/10.1016/j.physletb.2009.12.006} {\bibfield  {journal}
  {\bibinfo  {journal} {Phys. Lett. B}\ }\textbf {\bibinfo {volume} {683}},\
  \bibinfo {pages} {140} (\bibinfo {year} {2010})}\BibitemShut {NoStop}%
\bibitem [{\citenamefont {Xie}\ \emph {et~al.}(2013)\citenamefont {Xie},
  \citenamefont {Su}, \citenamefont {Zhu},\ and\ \citenamefont
  {Zhang}}]{Xie:2013np}%
  \BibitemOpen
  \bibfield  {author} {\bibinfo {author} {\bibfnamefont {W.-J.}\ \bibnamefont
  {Xie}}, \bibinfo {author} {\bibfnamefont {J.}~\bibnamefont {Su}}, \bibinfo
  {author} {\bibfnamefont {L.}~\bibnamefont {Zhu}},\ and\ \bibinfo {author}
  {\bibfnamefont {F.-S.}\ \bibnamefont {Zhang}},\ }\href
  {https://doi.org/10.1016/j.physletb.2012.12.021} {\bibfield  {journal}
  {\bibinfo  {journal} {Phys. Lett. B}\ }\textbf {\bibinfo {volume} {718}},\
  \bibinfo {pages} {1510} (\bibinfo {year} {2013})}\BibitemShut {NoStop}%
\bibitem [{\citenamefont {Hong}\ and\ \citenamefont
  {Danielewicz}(2014)}]{Hong:2013yva}%
  \BibitemOpen
  \bibfield  {author} {\bibinfo {author} {\bibfnamefont {J.}~\bibnamefont
  {Hong}}\ and\ \bibinfo {author} {\bibfnamefont {P.}~\bibnamefont
  {Danielewicz}},\ }\href {https://doi.org/10.1103/PhysRevC.90.024605}
  {\bibfield  {journal} {\bibinfo  {journal} {Phys. Rev. C}\ }\textbf {\bibinfo
  {volume} {90}},\ \bibinfo {pages} {024605} (\bibinfo {year}
  {2014})}\BibitemShut {NoStop}%
\bibitem [{\citenamefont {Cozma}(2017)}]{Cozma:2016qej}%
  \BibitemOpen
  \bibfield  {author} {\bibinfo {author} {\bibfnamefont {M.~D.}\ \bibnamefont
  {Cozma}},\ }\href {https://doi.org/10.1103/PhysRevC.95.014601} {\bibfield
  {journal} {\bibinfo  {journal} {Phys. Rev. C}\ }\textbf {\bibinfo {volume}
  {95}},\ \bibinfo {pages} {014601} (\bibinfo {year} {2017})}\BibitemShut
  {NoStop}%
\bibitem [{\citenamefont {Zhang}\ and\ \citenamefont
  {Ko}(2017)}]{Zhang:2017mps}%
  \BibitemOpen
  \bibfield  {author} {\bibinfo {author} {\bibfnamefont {Z.}~\bibnamefont
  {Zhang}}\ and\ \bibinfo {author} {\bibfnamefont {C.~M.}\ \bibnamefont {Ko}},\
  }\href {https://doi.org/10.1103/PhysRevC.95.064604} {\bibfield  {journal}
  {\bibinfo  {journal} {Phys. Rev. C}\ }\textbf {\bibinfo {volume} {95}},\
  \bibinfo {pages} {064604} (\bibinfo {year} {2017})}\BibitemShut {NoStop}%
\bibitem [{\citenamefont {Jhang}\ \emph {et~al.}(2021)\citenamefont {Jhang}
  \emph {et~al.}}]{SpiRIT:2020sfn}%
  \BibitemOpen
  \bibfield  {author} {\bibinfo {author} {\bibfnamefont {G.}~\bibnamefont
  {Jhang}} \emph {et~al.} (\bibinfo {collaboration} {SpiRIT, TMEP}),\ }\href
  {https://doi.org/10.1016/j.physletb.2020.136016} {\bibfield  {journal}
  {\bibinfo  {journal} {Phys. Lett. B}\ }\textbf {\bibinfo {volume} {813}},\
  \bibinfo {pages} {136016} (\bibinfo {year} {2021})}\BibitemShut {NoStop}%
\bibitem [{\citenamefont {Yong}(2021)}]{Yong:2021nwn}%
  \BibitemOpen
  \bibfield  {author} {\bibinfo {author} {\bibfnamefont {G.-C.}\ \bibnamefont
  {Yong}},\ }\href {https://doi.org/10.1103/PhysRevC.104.014613} {\bibfield
  {journal} {\bibinfo  {journal} {Phys. Rev. C}\ }\textbf {\bibinfo {volume}
  {104}},\ \bibinfo {pages} {014613} (\bibinfo {year} {2021})}\BibitemShut
  {NoStop}%
\bibitem [{\citenamefont {Chen}\ \emph {et~al.}(2007)\citenamefont {Chen},
  \citenamefont {Ko},\ and\ \citenamefont {Li}}]{Chen:2007}%
  \BibitemOpen
  \bibfield  {author} {\bibinfo {author} {\bibfnamefont {L.-W.}\ \bibnamefont
  {Chen}}, \bibinfo {author} {\bibfnamefont {C.~M.}\ \bibnamefont {Ko}},\ and\
  \bibinfo {author} {\bibfnamefont {B.-A.}\ \bibnamefont {Li}},\ }\href
  {https://doi.org/10.1103/PhysRevC.76.054316} {\bibfield  {journal} {\bibinfo
  {journal} {Phys. Rev. C}\ }\textbf {\bibinfo {volume} {76}},\ \bibinfo
  {pages} {054316} (\bibinfo {year} {2007})}\BibitemShut {NoStop}%
\bibitem [{\citenamefont {Chen}\ \emph {et~al.}(2009)\citenamefont {Chen},
  \citenamefont {Cai}, \citenamefont {Ko}, \citenamefont {Li}, \citenamefont
  {Shen},\ and\ \citenamefont {Xu}}]{Chen:2009wv}%
  \BibitemOpen
  \bibfield  {author} {\bibinfo {author} {\bibfnamefont {L.-W.}\ \bibnamefont
  {Chen}}, \bibinfo {author} {\bibfnamefont {B.-J.}\ \bibnamefont {Cai}},
  \bibinfo {author} {\bibfnamefont {C.~M.}\ \bibnamefont {Ko}}, \bibinfo
  {author} {\bibfnamefont {B.-A.}\ \bibnamefont {Li}}, \bibinfo {author}
  {\bibfnamefont {C.}~\bibnamefont {Shen}},\ and\ \bibinfo {author}
  {\bibfnamefont {J.}~\bibnamefont {Xu}},\ }\href
  {https://doi.org/10.1103/PhysRevC.80.014322} {\bibfield  {journal} {\bibinfo
  {journal} {Phys. Rev. C}\ }\textbf {\bibinfo {volume} {80}},\ \bibinfo
  {pages} {014322} (\bibinfo {year} {2009})}\BibitemShut {NoStop}%
\bibitem [{\citenamefont {Dutra}\ \emph {et~al.}(2012)\citenamefont {Dutra},
  \citenamefont {Lourenco}, \citenamefont {Sa~Martins}, \citenamefont
  {Delfino}, \citenamefont {Stone},\ and\ \citenamefont
  {Stevenson}}]{Dutra:2012mb}%
  \BibitemOpen
  \bibfield  {author} {\bibinfo {author} {\bibfnamefont {M.}~\bibnamefont
  {Dutra}}, \bibinfo {author} {\bibfnamefont {O.}~\bibnamefont {Lourenco}},
  \bibinfo {author} {\bibfnamefont {J.~S.}\ \bibnamefont {Sa~Martins}},
  \bibinfo {author} {\bibfnamefont {A.}~\bibnamefont {Delfino}}, \bibinfo
  {author} {\bibfnamefont {J.~R.}\ \bibnamefont {Stone}},\ and\ \bibinfo
  {author} {\bibfnamefont {P.~D.}\ \bibnamefont {Stevenson}},\ }\href
  {https://doi.org/10.1103/PhysRevC.85.035201} {\bibfield  {journal} {\bibinfo
  {journal} {Phys. Rev. C}\ }\textbf {\bibinfo {volume} {85}},\ \bibinfo
  {pages} {035201} (\bibinfo {year} {2012})}\BibitemShut {NoStop}%
\bibitem [{\citenamefont {Dutra}\ \emph {et~al.}(2014)\citenamefont {Dutra},
  \citenamefont {Louren\c{c}o}, \citenamefont {Avancini}, \citenamefont
  {Carlson}, \citenamefont {Delfino}, \citenamefont {Menezes}, \citenamefont
  {Provid\^encia}, \citenamefont {Typel},\ and\ \citenamefont
  {Stone}}]{Dutra:2014qga}%
  \BibitemOpen
  \bibfield  {author} {\bibinfo {author} {\bibfnamefont {M.}~\bibnamefont
  {Dutra}}, \bibinfo {author} {\bibfnamefont {O.}~\bibnamefont {Louren\c{c}o}},
  \bibinfo {author} {\bibfnamefont {S.~S.}\ \bibnamefont {Avancini}}, \bibinfo
  {author} {\bibfnamefont {B.~V.}\ \bibnamefont {Carlson}}, \bibinfo {author}
  {\bibfnamefont {A.}~\bibnamefont {Delfino}}, \bibinfo {author} {\bibfnamefont
  {D.~P.}\ \bibnamefont {Menezes}}, \bibinfo {author} {\bibfnamefont
  {C.}~\bibnamefont {Provid\^encia}}, \bibinfo {author} {\bibfnamefont
  {S.}~\bibnamefont {Typel}},\ and\ \bibinfo {author} {\bibfnamefont {J.~R.}\
  \bibnamefont {Stone}},\ }\href {https://doi.org/10.1103/PhysRevC.90.055203}
  {\bibfield  {journal} {\bibinfo  {journal} {Phys. Rev. C}\ }\textbf {\bibinfo
  {volume} {90}},\ \bibinfo {pages} {055203} (\bibinfo {year}
  {2014})}\BibitemShut {NoStop}%
\bibitem [{\citenamefont {Sun}\ \emph {et~al.}(2023)\citenamefont {Sun},
  \citenamefont {Bhattiprolu},\ and\ \citenamefont {Lattimer}}]{Sun:2023xkg}%
  \BibitemOpen
  \bibfield  {author} {\bibinfo {author} {\bibfnamefont {B.}~\bibnamefont
  {Sun}}, \bibinfo {author} {\bibfnamefont {S.}~\bibnamefont {Bhattiprolu}},\
  and\ \bibinfo {author} {\bibfnamefont {J.~M.}\ \bibnamefont {Lattimer}},\
  }\href@noop {} {\bibinfo {title} {{Compiled Properties of Nucleonic Matter
  and Nuclear and Neutron Star Models from Non-Relativistic and Relativistic
  Interactions}}} (\bibinfo {year} {2023}),\ \Eprint
  {https://arxiv.org/abs/2311.00843} {arXiv:2311.00843 [nucl-th]} \BibitemShut
  {NoStop}%
\bibitem [{\citenamefont {von Toussaint}(2011)}]{RevModPhys.83.943}%
  \BibitemOpen
  \bibfield  {author} {\bibinfo {author} {\bibfnamefont {U.}~\bibnamefont {von
  Toussaint}},\ }\href {https://doi.org/10.1103/RevModPhys.83.943} {\bibfield
  {journal} {\bibinfo  {journal} {Rev. Mod. Phys.}\ }\textbf {\bibinfo {volume}
  {83}},\ \bibinfo {pages} {943} (\bibinfo {year} {2011})}\BibitemShut
  {NoStop}%
\bibitem [{\citenamefont {Kejzlar}\ \emph {et~al.}(2020)\citenamefont
  {Kejzlar}, \citenamefont {Neufcourt}, \citenamefont {Nazarewicz},\ and\
  \citenamefont {Reinhard}}]{Kejzlar_2020}%
  \BibitemOpen
  \bibfield  {author} {\bibinfo {author} {\bibfnamefont {V.}~\bibnamefont
  {Kejzlar}}, \bibinfo {author} {\bibfnamefont {L.}~\bibnamefont {Neufcourt}},
  \bibinfo {author} {\bibfnamefont {W.}~\bibnamefont {Nazarewicz}},\ and\
  \bibinfo {author} {\bibfnamefont {P.-G.}\ \bibnamefont {Reinhard}},\ }\href
  {https://doi.org/10.1088/1361-6471/ab907c} {\bibfield  {journal} {\bibinfo
  {journal} {Journal of Physics G: Nuclear and Particle Physics}\ }\textbf
  {\bibinfo {volume} {47}},\ \bibinfo {pages} {094001} (\bibinfo {year}
  {2020})}\BibitemShut {NoStop}%
\bibitem [{\citenamefont {Jay}\ and\ \citenamefont {Neil}(2021)}]{Jay:2020jkz}%
  \BibitemOpen
  \bibfield  {author} {\bibinfo {author} {\bibfnamefont {W.~I.}\ \bibnamefont
  {Jay}}\ and\ \bibinfo {author} {\bibfnamefont {E.~T.}\ \bibnamefont {Neil}},\
  }\href {https://doi.org/10.1103/PhysRevD.103.114502} {\bibfield  {journal}
  {\bibinfo  {journal} {Phys. Rev. D}\ }\textbf {\bibinfo {volume} {103}},\
  \bibinfo {pages} {114502} (\bibinfo {year} {2021})}\BibitemShut {NoStop}%
\bibitem [{\citenamefont {Neufcourt}\ \emph {et~al.}(2019)\citenamefont
  {Neufcourt}, \citenamefont {Cao}, \citenamefont {Nazarewicz}, \citenamefont
  {Olsen},\ and\ \citenamefont {Viens}}]{Neufcourt:2019qvd}%
  \BibitemOpen
  \bibfield  {author} {\bibinfo {author} {\bibfnamefont {L.}~\bibnamefont
  {Neufcourt}}, \bibinfo {author} {\bibfnamefont {Y.}~\bibnamefont {Cao}},
  \bibinfo {author} {\bibfnamefont {W.}~\bibnamefont {Nazarewicz}}, \bibinfo
  {author} {\bibfnamefont {E.}~\bibnamefont {Olsen}},\ and\ \bibinfo {author}
  {\bibfnamefont {F.}~\bibnamefont {Viens}},\ }\href
  {https://doi.org/10.1103/PhysRevLett.122.062502} {\bibfield  {journal}
  {\bibinfo  {journal} {Phys. Rev. Lett.}\ }\textbf {\bibinfo {volume} {122}},\
  \bibinfo {pages} {062502} (\bibinfo {year} {2019})}\BibitemShut {NoStop}%
\bibitem [{\citenamefont {Everett}\ \emph {et~al.}(2021)\citenamefont {Everett}
  \emph {et~al.}}]{JETSCAPE:2020shq}%
  \BibitemOpen
  \bibfield  {author} {\bibinfo {author} {\bibfnamefont {D.}~\bibnamefont
  {Everett}} \emph {et~al.} (\bibinfo {collaboration} {JETSCAPE}),\ }\href
  {https://doi.org/10.1103/PhysRevLett.126.242301} {\bibfield  {journal}
  {\bibinfo  {journal} {Phys. Rev. Lett.}\ }\textbf {\bibinfo {volume} {126}},\
  \bibinfo {pages} {242301} (\bibinfo {year} {2021})}\BibitemShut {NoStop}%
\bibitem [{\citenamefont {Cirigliano}\ \emph {et~al.}(2022)\citenamefont
  {Cirigliano} \emph {et~al.}}]{Cirigliano:2022rmf}%
  \BibitemOpen
  \bibfield  {author} {\bibinfo {author} {\bibfnamefont {V.}~\bibnamefont
  {Cirigliano}} \emph {et~al.},\ }\href
  {https://doi.org/10.1088/1361-6471/aca03e} {\bibfield  {journal} {\bibinfo
  {journal} {J. Phys. G}\ }\textbf {\bibinfo {volume} {49}},\ \bibinfo {pages}
  {120502} (\bibinfo {year} {2022})}\BibitemShut {NoStop}%
\bibitem [{\citenamefont {Liu}\ \emph {et~al.}(2011)\citenamefont {Liu},
  \citenamefont {Wang}, \citenamefont {Deng},\ and\ \citenamefont
  {Wu}}]{Wangning:2011}%
  \BibitemOpen
  \bibfield  {author} {\bibinfo {author} {\bibfnamefont {M.}~\bibnamefont
  {Liu}}, \bibinfo {author} {\bibfnamefont {N.}~\bibnamefont {Wang}}, \bibinfo
  {author} {\bibfnamefont {Y.}~\bibnamefont {Deng}},\ and\ \bibinfo {author}
  {\bibfnamefont {X.}~\bibnamefont {Wu}},\ }\href
  {https://doi.org/10.1103/PhysRevC.84.014333} {\bibfield  {journal} {\bibinfo
  {journal} {Phys. Rev. C}\ }\textbf {\bibinfo {volume} {84}},\ \bibinfo
  {pages} {014333} (\bibinfo {year} {2011})}\BibitemShut {NoStop}%
\bibitem [{\citenamefont {Centelles}\ \emph {et~al.}(2009)\citenamefont
  {Centelles}, \citenamefont {Roca-Maza}, \citenamefont {Vi\~nas},\ and\
  \citenamefont {Warda}}]{Centelles:2009}%
  \BibitemOpen
  \bibfield  {author} {\bibinfo {author} {\bibfnamefont {M.}~\bibnamefont
  {Centelles}}, \bibinfo {author} {\bibfnamefont {X.}~\bibnamefont
  {Roca-Maza}}, \bibinfo {author} {\bibfnamefont {X.}~\bibnamefont {Vi\~nas}},\
  and\ \bibinfo {author} {\bibfnamefont {M.}~\bibnamefont {Warda}},\ }\href
  {https://doi.org/10.1103/PhysRevLett.102.122502} {\bibfield  {journal}
  {\bibinfo  {journal} {Phys. Rev. Lett.}\ }\textbf {\bibinfo {volume} {102}},\
  \bibinfo {pages} {122502} (\bibinfo {year} {2009})}\BibitemShut {NoStop}%
\bibitem [{\citenamefont {Chen}(2011)}]{Chen:2011}%
  \BibitemOpen
  \bibfield  {author} {\bibinfo {author} {\bibfnamefont {L.-W.}\ \bibnamefont
  {Chen}},\ }\href {https://doi.org/10.1103/PhysRevC.83.044308} {\bibfield
  {journal} {\bibinfo  {journal} {Phys. Rev. C}\ }\textbf {\bibinfo {volume}
  {83}},\ \bibinfo {pages} {044308} (\bibinfo {year} {2011})}\BibitemShut
  {NoStop}%
\bibitem [{\citenamefont {Wang}\ \emph {et~al.}(2013)\citenamefont {Wang},
  \citenamefont {Ou},\ and\ \citenamefont {Liu}}]{Wang:2013}%
  \BibitemOpen
  \bibfield  {author} {\bibinfo {author} {\bibfnamefont {N.}~\bibnamefont
  {Wang}}, \bibinfo {author} {\bibfnamefont {L.}~\bibnamefont {Ou}},\ and\
  \bibinfo {author} {\bibfnamefont {M.}~\bibnamefont {Liu}},\ }\href
  {https://doi.org/10.1103/PhysRevC.87.034327} {\bibfield  {journal} {\bibinfo
  {journal} {Phys. Rev. C}\ }\textbf {\bibinfo {volume} {87}},\ \bibinfo
  {pages} {034327} (\bibinfo {year} {2013})}\BibitemShut {NoStop}%
\bibitem [{\citenamefont {Chabanat}\ \emph {et~al.}(1997)\citenamefont
  {Chabanat}, \citenamefont {Meyer}, \citenamefont {Bonche}, \citenamefont
  {Schaeffer},\ and\ \citenamefont {Haensel}}]{Chabanat:1997qh}%
  \BibitemOpen
  \bibfield  {author} {\bibinfo {author} {\bibfnamefont {E.}~\bibnamefont
  {Chabanat}}, \bibinfo {author} {\bibfnamefont {J.}~\bibnamefont {Meyer}},
  \bibinfo {author} {\bibfnamefont {P.}~\bibnamefont {Bonche}}, \bibinfo
  {author} {\bibfnamefont {R.}~\bibnamefont {Schaeffer}},\ and\ \bibinfo
  {author} {\bibfnamefont {P.}~\bibnamefont {Haensel}},\ }\href
  {https://doi.org/10.1016/S0375-9474(97)00596-4} {\bibfield  {journal}
  {\bibinfo  {journal} {Nucl. Phys. A}\ }\textbf {\bibinfo {volume} {627}},\
  \bibinfo {pages} {710} (\bibinfo {year} {1997})}\BibitemShut {NoStop}%
\bibitem [{\citenamefont {Chen}\ \emph {et~al.}(2010)\citenamefont {Chen},
  \citenamefont {Ko}, \citenamefont {Li},\ and\ \citenamefont
  {Xu}}]{Chen:2010qx}%
  \BibitemOpen
  \bibfield  {author} {\bibinfo {author} {\bibfnamefont {L.-W.}\ \bibnamefont
  {Chen}}, \bibinfo {author} {\bibfnamefont {C.~M.}\ \bibnamefont {Ko}},
  \bibinfo {author} {\bibfnamefont {B.-A.}\ \bibnamefont {Li}},\ and\ \bibinfo
  {author} {\bibfnamefont {J.}~\bibnamefont {Xu}},\ }\href
  {https://doi.org/10.1103/PhysRevC.82.024321} {\bibfield  {journal} {\bibinfo
  {journal} {Phys. Rev. C}\ }\textbf {\bibinfo {volume} {82}},\ \bibinfo
  {pages} {024321} (\bibinfo {year} {2010})}\BibitemShut {NoStop}%
\bibitem [{\citenamefont {Kortelainen}\ \emph {et~al.}(2010)\citenamefont
  {Kortelainen}, \citenamefont {Lesinski}, \citenamefont {Mor{\'{e}}},
  \citenamefont {Nazarewicz}, \citenamefont {Sarich}, \citenamefont {Schunck},
  \citenamefont {Stoitsov},\ and\ \citenamefont {Wild}}]{Kortelainen2010}%
  \BibitemOpen
  \bibfield  {author} {\bibinfo {author} {\bibfnamefont {M.}~\bibnamefont
  {Kortelainen}}, \bibinfo {author} {\bibfnamefont {T.}~\bibnamefont
  {Lesinski}}, \bibinfo {author} {\bibfnamefont {J.}~\bibnamefont
  {Mor{\'{e}}}}, \bibinfo {author} {\bibfnamefont {W.}~\bibnamefont
  {Nazarewicz}}, \bibinfo {author} {\bibfnamefont {J.}~\bibnamefont {Sarich}},
  \bibinfo {author} {\bibfnamefont {N.}~\bibnamefont {Schunck}}, \bibinfo
  {author} {\bibfnamefont {M.~V.}\ \bibnamefont {Stoitsov}},\ and\ \bibinfo
  {author} {\bibfnamefont {S.}~\bibnamefont {Wild}},\ }\href
  {https://doi.org/10.1103/PhysRevC.82.024313} {\bibfield  {journal} {\bibinfo
  {journal} {Phys. Rev. C}\ }\textbf {\bibinfo {volume} {82}},\ \bibinfo
  {pages} {024313} (\bibinfo {year} {2010})}\BibitemShut {NoStop}%
\bibitem [{\citenamefont {Chen}\ and\ \citenamefont {Gu}(2012)}]{Chen:2011ps}%
  \BibitemOpen
  \bibfield  {author} {\bibinfo {author} {\bibfnamefont {L.-W.}\ \bibnamefont
  {Chen}}\ and\ \bibinfo {author} {\bibfnamefont {J.-Z.}\ \bibnamefont {Gu}},\
  }\href {https://doi.org/10.1088/0954-3899/39/3/035104} {\bibfield  {journal}
  {\bibinfo  {journal} {J. Phys. G}\ }\textbf {\bibinfo {volume} {39}},\
  \bibinfo {pages} {035104} (\bibinfo {year} {2012})}\BibitemShut {NoStop}%
\bibitem [{\citenamefont {Reinhard}(1991)}]{Reinhard1991}%
  \BibitemOpen
  \bibfield  {author} {\bibinfo {author} {\bibfnamefont {P.-G.}\ \bibnamefont
  {Reinhard}},\ }\bibinfo {title} {The skyrme---hartree---fock model of the
  nuclear ground state},\ in\ \href
  {https://doi.org/10.1007/978-3-642-76356-4_2} {\emph {\bibinfo {booktitle}
  {Computational Nuclear Physics 1: Nuclear Structure}}},\ \bibinfo {editor}
  {edited by\ \bibinfo {editor} {\bibfnamefont {K.}~\bibnamefont {Langanke}},
  \bibinfo {editor} {\bibfnamefont {J.~A.}\ \bibnamefont {Maruhn}},\ and\
  \bibinfo {editor} {\bibfnamefont {S.~E.}\ \bibnamefont {Koonin}}}\ (\bibinfo
  {publisher} {Springer Berlin Heidelberg},\ \bibinfo {address} {Berlin,
  Heidelberg},\ \bibinfo {year} {1991})\ pp.\ \bibinfo {pages}
  {28--50}\BibitemShut {NoStop}%
\bibitem [{\citenamefont {Wang}\ \emph {et~al.}(2021)\citenamefont {Wang},
  \citenamefont {Huang}, \citenamefont {Kondev}, \citenamefont {Audi},\ and\
  \citenamefont {Naimi}}]{AME2020}%
  \BibitemOpen
  \bibfield  {author} {\bibinfo {author} {\bibfnamefont {M.}~\bibnamefont
  {Wang}}, \bibinfo {author} {\bibfnamefont {W.}~\bibnamefont {Huang}},
  \bibinfo {author} {\bibfnamefont {F.}~\bibnamefont {Kondev}}, \bibinfo
  {author} {\bibfnamefont {G.}~\bibnamefont {Audi}},\ and\ \bibinfo {author}
  {\bibfnamefont {S.}~\bibnamefont {Naimi}},\ }\href
  {https://doi.org/10.1088/1674-1137/abddaf} {\bibfield  {journal} {\bibinfo
  {journal} {Chinese Physics C}\ }\textbf {\bibinfo {volume} {45}},\ \bibinfo
  {pages} {030003} (\bibinfo {year} {2021})}\BibitemShut {NoStop}%
\bibitem [{\citenamefont {Horowitz}\ and\ \citenamefont
  {Piekarewicz}(2001)}]{Horowitz:2000xj}%
  \BibitemOpen
  \bibfield  {author} {\bibinfo {author} {\bibfnamefont {C.~J.}\ \bibnamefont
  {Horowitz}}\ and\ \bibinfo {author} {\bibfnamefont {J.}~\bibnamefont
  {Piekarewicz}},\ }\href {https://doi.org/10.1103/PhysRevLett.86.5647}
  {\bibfield  {journal} {\bibinfo  {journal} {Phys. Rev. Lett.}\ }\textbf
  {\bibinfo {volume} {86}},\ \bibinfo {pages} {5647} (\bibinfo {year}
  {2001})}\BibitemShut {NoStop}%
\bibitem [{\citenamefont {Cai}\ and\ \citenamefont {Chen}(2017)}]{Cai:2014kya}%
  \BibitemOpen
  \bibfield  {author} {\bibinfo {author} {\bibfnamefont {B.-J.}\ \bibnamefont
  {Cai}}\ and\ \bibinfo {author} {\bibfnamefont {L.-W.}\ \bibnamefont {Chen}},\
  }\href {https://doi.org/10.1007/s41365-017-0329-1} {\bibfield  {journal}
  {\bibinfo  {journal} {Nucl. Sci. Tech.}\ }\textbf {\bibinfo {volume} {28}},\
  \bibinfo {pages} {185} (\bibinfo {year} {2017})}\BibitemShut {NoStop}%
\bibitem [{\citenamefont {Chen}\ and\ \citenamefont
  {Piekarewicz}(2014)}]{Chen:2014}%
  \BibitemOpen
  \bibfield  {author} {\bibinfo {author} {\bibfnamefont {W.-C.}\ \bibnamefont
  {Chen}}\ and\ \bibinfo {author} {\bibfnamefont {J.}~\bibnamefont
  {Piekarewicz}},\ }\href {https://doi.org/10.1103/PhysRevC.90.044305}
  {\bibfield  {journal} {\bibinfo  {journal} {Phys. Rev. C}\ }\textbf {\bibinfo
  {volume} {90}},\ \bibinfo {pages} {044305} (\bibinfo {year}
  {2014})}\BibitemShut {NoStop}%
\bibitem [{\citenamefont {Niksic}\ \emph {et~al.}(2014)\citenamefont {Niksic},
  \citenamefont {Paar}, \citenamefont {Vretenar},\ and\ \citenamefont
  {Ring}}]{Niksic:2014dra}%
  \BibitemOpen
  \bibfield  {author} {\bibinfo {author} {\bibfnamefont {T.}~\bibnamefont
  {Niksic}}, \bibinfo {author} {\bibfnamefont {N.}~\bibnamefont {Paar}},
  \bibinfo {author} {\bibfnamefont {D.}~\bibnamefont {Vretenar}},\ and\
  \bibinfo {author} {\bibfnamefont {P.}~\bibnamefont {Ring}},\ }\href
  {https://doi.org/10.1016/j.cpc.2014.02.027} {\bibfield  {journal} {\bibinfo
  {journal} {Comput. Phys. Commun.}\ }\textbf {\bibinfo {volume} {185}},\
  \bibinfo {pages} {1808} (\bibinfo {year} {2014})}\BibitemShut {NoStop}%
\bibitem [{\citenamefont {Plumlee}\ \emph {et~al.}(2023)\citenamefont
  {Plumlee}, \citenamefont {S\"urer}, \citenamefont {Wild},\ and\ \citenamefont
  {Chan}}]{surmise2023}%
  \BibitemOpen
  \bibfield  {author} {\bibinfo {author} {\bibfnamefont {M.}~\bibnamefont
  {Plumlee}}, \bibinfo {author} {\bibfnamefont {O.}~\bibnamefont {S\"urer}},
  \bibinfo {author} {\bibfnamefont {S.~M.}\ \bibnamefont {Wild}},\ and\
  \bibinfo {author} {\bibfnamefont {M.~Y.-H.}\ \bibnamefont {Chan}},\ }\href
  {https://surmise.readthedocs.io} {\emph {\bibinfo {title} {{surmise 0.2.0}
  Users Manual}}},\ \bibinfo {type} {Tech. Rep.}\ \bibinfo {number} {Version
  0.2.0}\ (\bibinfo  {institution} {NAISE},\ \bibinfo {year}
  {2023})\BibitemShut {NoStop}%
\bibitem [{\citenamefont {Bernhard}\ \emph {et~al.}(2015)\citenamefont
  {Bernhard}, \citenamefont {Marcy}, \citenamefont {Coleman-Smith},
  \citenamefont {Huzurbazar}, \citenamefont {Wolpert},\ and\ \citenamefont
  {Bass}}]{Bernhard:2015hxa}%
  \BibitemOpen
  \bibfield  {author} {\bibinfo {author} {\bibfnamefont {J.~E.}\ \bibnamefont
  {Bernhard}}, \bibinfo {author} {\bibfnamefont {P.~W.}\ \bibnamefont {Marcy}},
  \bibinfo {author} {\bibfnamefont {C.~E.}\ \bibnamefont {Coleman-Smith}},
  \bibinfo {author} {\bibfnamefont {S.}~\bibnamefont {Huzurbazar}}, \bibinfo
  {author} {\bibfnamefont {R.~L.}\ \bibnamefont {Wolpert}},\ and\ \bibinfo
  {author} {\bibfnamefont {S.~A.}\ \bibnamefont {Bass}},\ }\href
  {https://doi.org/10.1103/PhysRevC.91.054910} {\bibfield  {journal} {\bibinfo
  {journal} {Phys. Rev. C}\ }\textbf {\bibinfo {volume} {91}},\ \bibinfo
  {pages} {054910} (\bibinfo {year} {2015})}\BibitemShut {NoStop}%
\bibitem [{\citenamefont {Abril-Pla}\ \emph {et~al.}(2023)\citenamefont
  {Abril-Pla}, \citenamefont {Andreani}, \citenamefont {Carroll}, \citenamefont
  {Dong}, \citenamefont {Fonnesbeck}, \citenamefont {Kochurov}, \citenamefont
  {Kumar}, \citenamefont {Lao}, \citenamefont {Luhmann}, \citenamefont {Martin}
  \emph {et~al.}}]{abril2023pymc}%
  \BibitemOpen
  \bibfield  {author} {\bibinfo {author} {\bibfnamefont {O.}~\bibnamefont
  {Abril-Pla}}, \bibinfo {author} {\bibfnamefont {V.}~\bibnamefont {Andreani}},
  \bibinfo {author} {\bibfnamefont {C.}~\bibnamefont {Carroll}}, \bibinfo
  {author} {\bibfnamefont {L.}~\bibnamefont {Dong}}, \bibinfo {author}
  {\bibfnamefont {C.~J.}\ \bibnamefont {Fonnesbeck}}, \bibinfo {author}
  {\bibfnamefont {M.}~\bibnamefont {Kochurov}}, \bibinfo {author}
  {\bibfnamefont {R.}~\bibnamefont {Kumar}}, \bibinfo {author} {\bibfnamefont
  {J.}~\bibnamefont {Lao}}, \bibinfo {author} {\bibfnamefont {C.~C.}\
  \bibnamefont {Luhmann}}, \bibinfo {author} {\bibfnamefont {O.~A.}\
  \bibnamefont {Martin}}, \emph {et~al.},\ }\href@noop {} {\bibfield  {journal}
  {\bibinfo  {journal} {PeerJ Computer Science}\ }\textbf {\bibinfo {volume}
  {9}},\ \bibinfo {pages} {e1516} (\bibinfo {year} {2023})}\BibitemShut
  {NoStop}%
\bibitem [{\citenamefont {Lynch}\ and\ \citenamefont
  {Tsang}(2022)}]{Lynch:2021xkq}%
  \BibitemOpen
  \bibfield  {author} {\bibinfo {author} {\bibfnamefont {W.~G.}\ \bibnamefont
  {Lynch}}\ and\ \bibinfo {author} {\bibfnamefont {M.~B.}\ \bibnamefont
  {Tsang}},\ }\href {https://doi.org/10.1016/j.physletb.2022.137098} {\bibfield
   {journal} {\bibinfo  {journal} {Phys. Lett. B}\ }\textbf {\bibinfo {volume}
  {830}},\ \bibinfo {pages} {137098} (\bibinfo {year} {2022})}\BibitemShut
  {NoStop}%
\bibitem [{\citenamefont {Drischler}\ \emph {et~al.}(2020)\citenamefont
  {Drischler}, \citenamefont {Furnstahl}, \citenamefont {Melendez},\ and\
  \citenamefont {Phillips}}]{Drischler:2020hwi}%
  \BibitemOpen
  \bibfield  {author} {\bibinfo {author} {\bibfnamefont {C.}~\bibnamefont
  {Drischler}}, \bibinfo {author} {\bibfnamefont {R.~J.}\ \bibnamefont
  {Furnstahl}}, \bibinfo {author} {\bibfnamefont {J.~A.}\ \bibnamefont
  {Melendez}},\ and\ \bibinfo {author} {\bibfnamefont {D.~R.}\ \bibnamefont
  {Phillips}},\ }\href {https://doi.org/10.1103/PhysRevLett.125.202702}
  {\bibfield  {journal} {\bibinfo  {journal} {Phys. Rev. Lett.}\ }\textbf
  {\bibinfo {volume} {125}},\ \bibinfo {pages} {202702} (\bibinfo {year}
  {2020})}\BibitemShut {NoStop}%
\bibitem [{\citenamefont {Brown}(2013)}]{Brown:2013mga}%
  \BibitemOpen
  \bibfield  {author} {\bibinfo {author} {\bibfnamefont {B.~A.}\ \bibnamefont
  {Brown}},\ }\href {https://doi.org/10.1103/PhysRevLett.111.232502} {\bibfield
   {journal} {\bibinfo  {journal} {Phys. Rev. Lett.}\ }\textbf {\bibinfo
  {volume} {111}},\ \bibinfo {pages} {232502} (\bibinfo {year}
  {2013})}\BibitemShut {NoStop}%
\bibitem [{\citenamefont {Trippa}\ \emph {et~al.}(2008)\citenamefont {Trippa},
  \citenamefont {Colo},\ and\ \citenamefont {Vigezzi}}]{Trippa:2008gr}%
  \BibitemOpen
  \bibfield  {author} {\bibinfo {author} {\bibfnamefont {L.}~\bibnamefont
  {Trippa}}, \bibinfo {author} {\bibfnamefont {G.}~\bibnamefont {Colo}},\ and\
  \bibinfo {author} {\bibfnamefont {E.}~\bibnamefont {Vigezzi}},\ }\href
  {https://doi.org/10.1103/PhysRevC.77.061304} {\bibfield  {journal} {\bibinfo
  {journal} {Phys. Rev. C}\ }\textbf {\bibinfo {volume} {77}},\ \bibinfo
  {pages} {061304} (\bibinfo {year} {2008})}\BibitemShut {NoStop}%
\bibitem [{\citenamefont {Zhang}\ and\ \citenamefont
  {Chen}(2013)}]{Zhang:2013wna}%
  \BibitemOpen
  \bibfield  {author} {\bibinfo {author} {\bibfnamefont {Z.}~\bibnamefont
  {Zhang}}\ and\ \bibinfo {author} {\bibfnamefont {L.-W.}\ \bibnamefont
  {Chen}},\ }\href {https://doi.org/10.1016/j.physletb.2013.08.002} {\bibfield
  {journal} {\bibinfo  {journal} {Phys. Lett. B}\ }\textbf {\bibinfo {volume}
  {726}},\ \bibinfo {pages} {234} (\bibinfo {year} {2013})}\BibitemShut
  {NoStop}%
\bibitem [{\citenamefont {Wang}\ and\ \citenamefont
  {Chen}(2015)}]{Wang:2014mra}%
  \BibitemOpen
  \bibfield  {author} {\bibinfo {author} {\bibfnamefont {R.}~\bibnamefont
  {Wang}}\ and\ \bibinfo {author} {\bibfnamefont {L.-W.}\ \bibnamefont
  {Chen}},\ }\href {https://doi.org/10.1103/PhysRevC.92.031303} {\bibfield
  {journal} {\bibinfo  {journal} {Phys. Rev. C}\ }\textbf {\bibinfo {volume}
  {92}},\ \bibinfo {pages} {031303} (\bibinfo {year} {2015})}\BibitemShut
  {NoStop}%
\end{thebibliography}%

\end{document}